\RequirePackage{fixltx2e}
\documentclass[aps,prl,reprint,showpacs,twocolumn,amsmath,amssymb,floatfix,nofootinbib,preprintnumbers]{revtex4-1}

\usepackage{graphicx}
\usepackage{dcolumn}
\usepackage{bm}
\usepackage{subfigure} 
\usepackage{amsmath}
\usepackage{amsfonts}
\usepackage{slashed}
\usepackage{hyperref}
\usepackage{multirow}
\usepackage{soul}
\usepackage{mathtools}
\usepackage{float}
\usepackage{afterpage}
\usepackage{color}
\usepackage{latexsym} 
\usepackage{rotating} 
\usepackage{verbatim} 
\usepackage{microtype} 
\usepackage{marginnote}
\usepackage[utf8]{inputenc}
\usepackage[outdir=./]{epstopdf}

\hypersetup{%
colorlinks = {true},
filecolor = {black},
linkcolor = {blue},
menucolor = {black},
citecolor = {blue},
urlcolor = {blue},
}{}

\begin{document}
\begin{flushright}
MSUHEP-17-006
\end{flushright}

\title{Variance Reduction and Cluster Decomposition}

\author{Keh-Fei Liu$^{1}$, Jian Liang$^{1}$, and Yi-Bo Yang$^{2}$
}
\affiliation{
$^{1}$\mbox{Department of Physics and Astronomy, University of Kentucky, Lexington, KY 40506, USA}
$^{2}$\mbox{Department of Physics and Astronomy, Michigan State University, East Lansing, MI 48824, USA}}

\begin{abstract}
It is a common problem in lattice QCD calculation of the mass of the hadron with an annihilation channel that the signal falls off in time while the noise remains constant. In addition, the disconnected insertion calculation of the three-point function and the calculation of the neutron electric dipole moment with the $\theta$ term suffer from a noise problem due to the $\sqrt{V}$ fluctuation.  We identify these problems to have the same origin and the $\sqrt{V}$ problem can be overcome by utilizing the cluster decomposition principle.  We demonstrate this by considering the calculations of the glueball mass, the strangeness content in the nucleon, and the CP violation angle in the nucleon due to the $\theta$ term. It is found that for lattices with physical sizes of 4.5 - 5.5 fm, the statistical errors of these quantities can be reduced by a factor of  3 to 4. 
The systematic errors can be estimated from the Akaike information criterion. For the strangeness content, we find that the systematic error is of
the same size as that of the statistical one when the cluster decomposition principle is utilized. This results in a 2 to 3 times reduction in the overall error.

\end{abstract}

\maketitle


\section{Introduction}
As the physical pion mass is accessible in lattice QCD simulations nowadays with larger physical volumes and several lattice spacings with different lattice actions to estimate the associated systematic errors, lattice QCD calculation is getting mature, particularly
for flavor physics where the quark masses, heavy-light decay constants, CKM matrices, and strong coupling constant are
reviewed and averaged by FLAG~\cite{Aoki:2016frl}.  On the other hand, the baryon physics is not as settled as that of
mesons. Part of the reasons is illustrated in the Parisi-Lepage consideration of the signal-to-noise ratio of the nucleon two-point function. Since the variance of the nucleon propagator has three-pions as the lowest state in the correlator, the signal-to-noise (S/N) ratio is proportional to $e^{-(m_N - 3/2 m_{\pi})t}$~\cite{Parisi:1983ae,Lepage:1989hd} and noticeably grows exponentially with $t$ when the pion mass is close to the physical one in lattice calculations. This is why baryon physics is more noisy than that of mesons.

One special problem associated with the correlators of mesons  involving annihilation channels or glueballs is that the signal falls off exponential with time, but the noise remains constant. Thus, after certain time separation, the signal falls below the noise and succumbs to the sign problem. Another aspect of the DI is observed in the DI three-point function involving a quark loop or the topological charge in the neutron electric dipole moment (nEDM) calculation from the $\theta$ term, the fluctuations of the quark loop and the topological charge are proportion to $\sqrt{V}$ which pose a challenge for calculations large volumes lattices.
In this work, we shall show that the constant error and $\sqrt{V}$ fluctuation in the DI have the same origin and they can be ameliorated with the help of the property of the cluster decomposition principle so that the S/N ratio can be improved by a factor of
$\sqrt{V/V_{R_s}}$ where $V_{R_s}$ is the volume with radius $R_s$ which is the effective correlation length between the operators.

\section{Cluster decomposition principle and variance reduction}
   One often invokes the locality argument to justify that experiments conducted on Earth is not affected by events on the Moon.
 This is a consequence of the cluster decomposition principle (CDP) in that if color-singlet operators in a correlator are separated by a large enough  space-like distance, the correlator will be zero. In other words, the operators are not correlated in this circumstance.
To be specific, it is shown~\cite{Araki:1962zhd} that under the assumptions of translation invariance, stability of the vacuum, existence of a lowest non-zero mass and local commutativity, one has 
\begin{equation}   \label{CDP}
\begin{split}
&|\langle0|\mathcal{B}_1(x_1)\mathcal{B}_2(x_2)|0\rangle_s |\leq A r^{-\frac{2}{3}}e^{-Mr}
\end{split}
\end{equation}
for a large enough space-like distance \mbox{$r=|x_1-x_2|$,} 
where $\langle0|\mathcal{B}_1(x_1)\mathcal{B}_2(x_2)|0\rangle_s\equiv\langle0|\mathcal{B}_1(x_1)\mathcal{B}_2(x_2)|0\rangle
-\langle 0|\mathcal{B}_1(x_1)|0\rangle \langle 0|\mathcal{B}_2(x_2)|0\rangle$ is the vacuum-subtracted correlation function.
 $\mathcal{B}_1(x_1)$ and $\mathcal{B}_2(x_2)$ are two color-singlet operator clusters whose centers are at $x_1$ and $x_2$ respectively, $M$ is the smallest non-zero mass for the correlator, and $A$ is a constant. This is the asymptotic behavior of a boson propagator $K_1(r)/r$. This means the correlation between two operator clusters far apart with large enough space-like distance $r$ tends to be zero at least as fast as $r^{-\frac{2}{3}}e^{-Mr}$. Given that the longest correlation length in QCD is $1/m_{\pi}$, one has $M \ge m_{\pi}$. Since the Euclidean separation is always `space-like', the cluster decomposition principle (CDP) is applicable to the Euclidean correlators. Some of the recent attempts to reduce variances in the calculations of strangeness in the nucleon~\cite{Bali:2009dz}, the $\rho$ meson mass~\cite{Chen:2015tpa}, the light-by-light contribution in the muonic $g-2$~\cite{Blum:2015gfa}, the factorization of fermion determinant~\cite{Ce:2016ajy}, and reweighting of nEDM calculation with topological charge density~\cite{Shintani:2015vsx} have applied concepts similar or related to that of CDP. In this work, we prove that, applying the CDP explicitly, the error of an DI correlator can be improved by a factor of $\sqrt{V/V_{R_s}}$.

In evaluating the correlators, one often takes a volume sum over the three-dimensional coordinates. To estimate at what distance the large distance behavior saturates, we integrate the fall-off to a cut off distance $R$,
\begin{eqnarray}
\label{integral}
\int_0^{R}\!\!\! d^3r\,r^{-\frac{3}{2}}e^{-Mr} 
\!= \!4\pi \!\left(\!\!\frac{\sqrt{\pi}{\rm erf}(\sqrt{MR})}{2M^{\frac{3}{2}}}\!-\!\frac{\sqrt{R}e^{-MR}}{M}\!\right)\!\!,
\end{eqnarray}
where ${\rm erf}$ is the error function. Since the kernel of the integral decays very quickly,  the integral has already gained more than $99.5\%$ of its total value for $R=8/M$. Assuming the fall-off behavior dominates the volume-integrated correlator, we
consider $R_s\sim\frac{8}{M}$ as an effective cutoff and the correlation with separation $r> R_s$ has negligible contribution.

To test the principle of cluster decomposition with lattice data, we consider the two-point correlator for a fixed $t$ 
with a cutoff of $R$ in the relative coordinate between the two color-singlet operators $\mathcal{O}_1$ and $\mathcal{O}_2$
\begin{equation}
\label{general_form}
C(R,t)=\frac{1}{V}\langle \sum_{\vec{x}}\sum_{r<R}\mathcal{O}_1(\vec{x}+\vec{r'},t)\, \mathcal{O}_2(\vec{x},0)\rangle,
\end{equation}
where $r = \sqrt{|\vec{r'}|^2 + t^2}$.
The correlation functions in the present work are calculated using valence overlap fermions 
on the RBC-UKQCD $2 + 1$ flavor domain-wall configurations.
More detailed definitions and numerical implementations can be found in previous works 
~\cite{Li:2010pw,Gong:2013vja,Sufian:2016pex,Yang:2015zja}.

We examine the nucleon two-point function first on the $48^3\times 96$ lattice (48I) with the physical sea quark 
mass~ \cite{Blum:2014tka}. We use 3 valence quark masses corresponding to pion masses $70$ MeV, $149$ MeV and $260$ MeV respectively.  

\begin{figure}[t]
\centering
\includegraphics[width=0.4\textwidth]{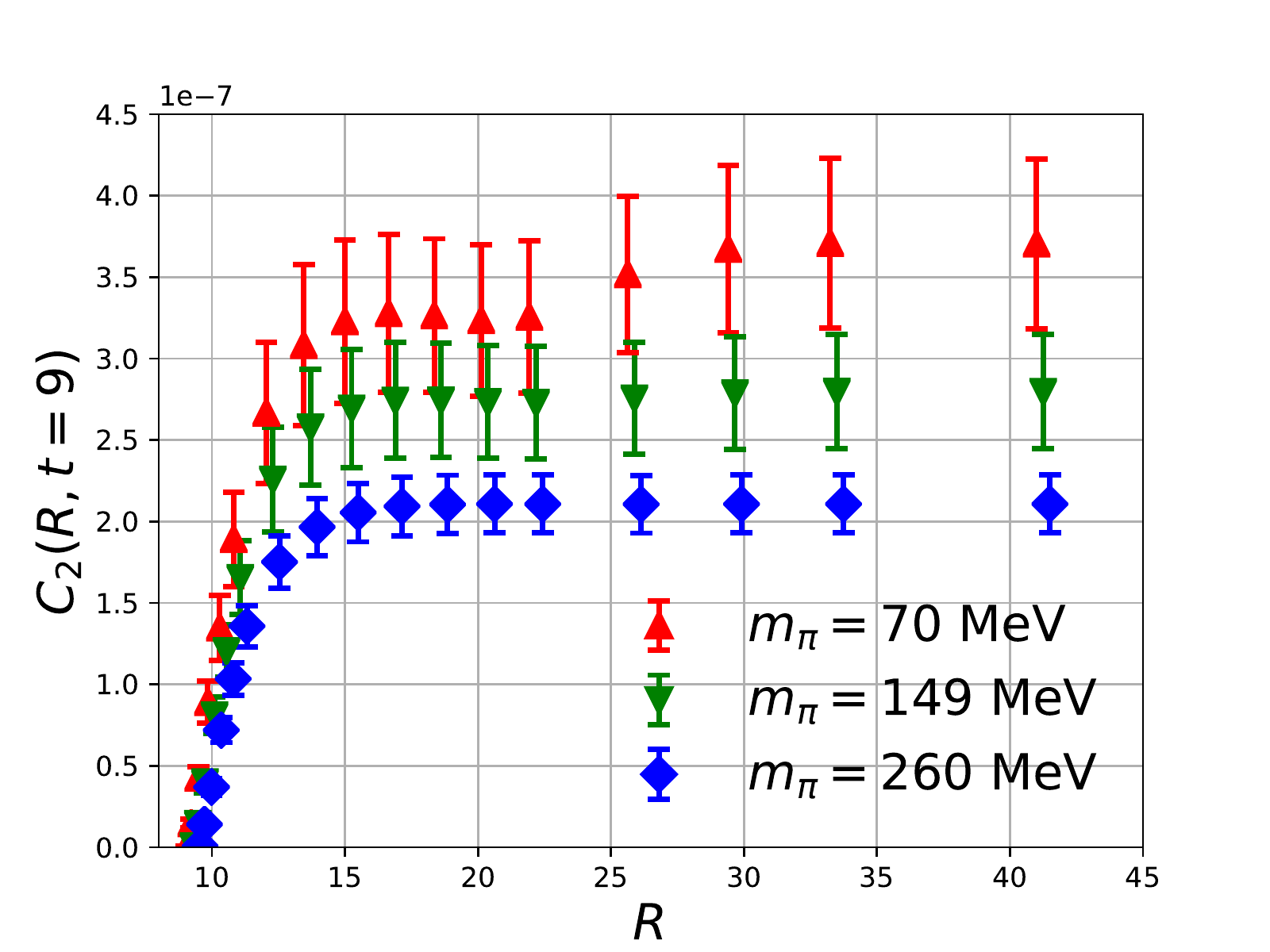}
\caption{Nucleon two-point functions at $t=9$ for three different valence quark masses as a function of the cutoff  $R$.}
\label{fig_twopoint}
\end{figure}

The results for the nucleon correlators at $t = 9$ for three different valence quark masses are plotted  in Fig.~\ref{fig_twopoint} as a function of $R$ which is the cutoff of the Euclidean distance $r$ between the point source and the sink. We see that the nucleon correlator basically saturates after \mbox{$R \sim 15 =  1.71$ fm} with $a = 0.114$ fm for the three cases. This agrees well with our earlier estimate of a saturation radius $R_s = 8/M$ which corresponds to $\sim 1.66$ fm. This shows that the CDP works and 
Eq.~(\ref{CDP}) gives a good estimate of $R_s$.

Since the signal of the correlators falls off exponentially with $r$, summing over $r$ beyond the saturating radius $R_s$ does not change the signal and will only gather noise. Let's consider the disconnected insertion next and see how the S/N ratio can be improved with this observation. In the case of the DI, the variance of the correlator in Eq.~(\ref{general_form}) 
\mbox{$\frac{1}{V^2}\langle |\sum_{\vec{x}}\sum_{r<R}\mathcal{O}_1(\vec{x}+\vec{r'},t)\, \mathcal{O}_2(\vec{x},0)|^2\rangle$} can have a vacuum insertion in addition to the exponential fall off in $t$ due to the $\mathcal{O}^{\dagger}\mathcal{O}$ operator. 
%
\begin{eqnarray}
\label{Variance}
V\!ar(R,t) &=&\frac{1}{V^2}\sum_{\vec{x},\vec{y}}\left(\langle \sum_{r_1 <R} \mathcal{O}_1(\vec{x}+\vec{r_1'},t)  \sum_{r_2 <R}\mathcal{O}_1^{\dagger}(\vec{y}+\vec{r'_2},t)\rangle\right.
 \nonumber \\
&\cdot& \left.\langle \mathcal{O}_2(\vec{x},0) \mathcal{O}_2^{\dagger}(\vec{y},0)\rangle\right) + ....,
\end{eqnarray}
where $r_1 = \sqrt{|\vec{r_1'}|^2 + t^2}$ and $r_2 = \sqrt{|\vec{r_2'}|^2 + t^2}$ respectively.
For the case where $\vec{r_1'}$ and $\vec{r_2'}$ are integrated over the whole lattice volume, 
the sum over the positions $\vec{x}, \vec{y}, \vec{x}+\vec{r'_1}$ and $\vec{y}+\vec{r'_2}$ can be carried out independently. Consequently, 
$\mathcal{O}_1$ and $\mathcal{O}_2$ in the DI fluctuate independently which leads to a variance which is the product of their respective variances. In this case, the leading vacuum insertion is a constant, independent of $t$. This is the reason why the noise remains constant over $t$ in DI.  On the other hand, the constant variance is reduced to $V_{R_s}/V$ when $r$ it is integrated to $R_s$, while the signal is not compromised. The sub-leading contribution (denoted by ...) in Eq.~(\ref{Variance}) has an exponential decay in $t$ with a mass in the scalar channel. It is clear that to leading order, the ratio of the cutoff S/N at $R_s$ to that without cutoff is
\begin{equation} \label{S/N_ratio}
\frac{S/N(R_s)}{S/N(L)} \sim \sqrt{\frac{V}{\,\,\,V_{R_s}}}.
\end{equation}

We shall consider several DI examples involving volume summations over two or more coordinates. Since the convoluted sum with a relative coordinate in Eq.~(\ref{general_form}) can be expensive, we shall invoke  the standard convolution theorem by calculating the product of two functions 
$\tilde{K}(\vec{p},t)=\tilde{\mathcal{O}}_1(-\vec{p})\tilde{\mathcal{O}}_2(\vec{p}),$ 
where $\tilde{\mathcal{O}}_1(-\vec{p})/\tilde{\mathcal{O}}_2(\vec{p})$ is the Fourier transforms of 
$\mathcal{O}_1 (\vec{x})/\mathcal{O}_2 (\vec{x})$ in each configuration on their respective time slices. Then
\begin{equation}   \label{general_form_integral_2}
C(R,t)=\langle \int_{r<R}d\vec{r'}K(\vec{r'},t)\rangle.
\end{equation}
where $K(\vec{r'},t)$ is the Fourier transform of $\tilde{K}(\vec{p},t)$.
In this way, the cost of the double-summation, which is order $V^2$, is reduced to  that of the fast Fourier transform (FFT)
which is in the order of $V\,{\rm log}V$. 

\subsection{Scalar matrix element of the strange quark}
The first example is the disconnected insertion for the nucleon matrix element with a scalar loop which involves a three-point function
and can be expressed as
\begin{equation}
\label{eq_dis_ins}
C_3(R,\tau,t)=\langle \sum_{\vec{x}}\sum_{r<R}\mathcal{O}_N(\vec{x},t) S(\vec{x}+\vec{r'},\tau)\bar{\mathcal{O}}_N(\mathcal{G},0)\rangle,
\end{equation}
where $S$ is the vacuum-subtracted scalar loop, $\mathcal{G}$ denotes the source grid for increasing statistics~\cite{Li:2010pw}. 
Note here $r = \sqrt{|\vec{r'}|^2 + (t-\tau)^2}$ is the 4-D distance and $r_x$ is the spatial separation between the loop and the sink.
Since the low-modes dominate the strangeness in the nucleon~\cite{Yang:2015uis}, we calculate the strange quark loop with
low-modes only to illustrate the CDP effect. The sum over the spatial relative coordinate between the scalar quark loop $S$ and the
sink interpolation operator $O_N(\vec{x},t)$ is carried out through the convolution in Eq.~(\ref{general_form_integral_2}).
This calculation is done on the domain-wall $32^3\times64$ (32ID) lattice \cite{Blum:2014tka} with pion 
mass $\sim170$ MeV and the lattice size is $4.6$ fm.

\begin{figure}[t]     
\centering
\includegraphics[width=0.4\textwidth]{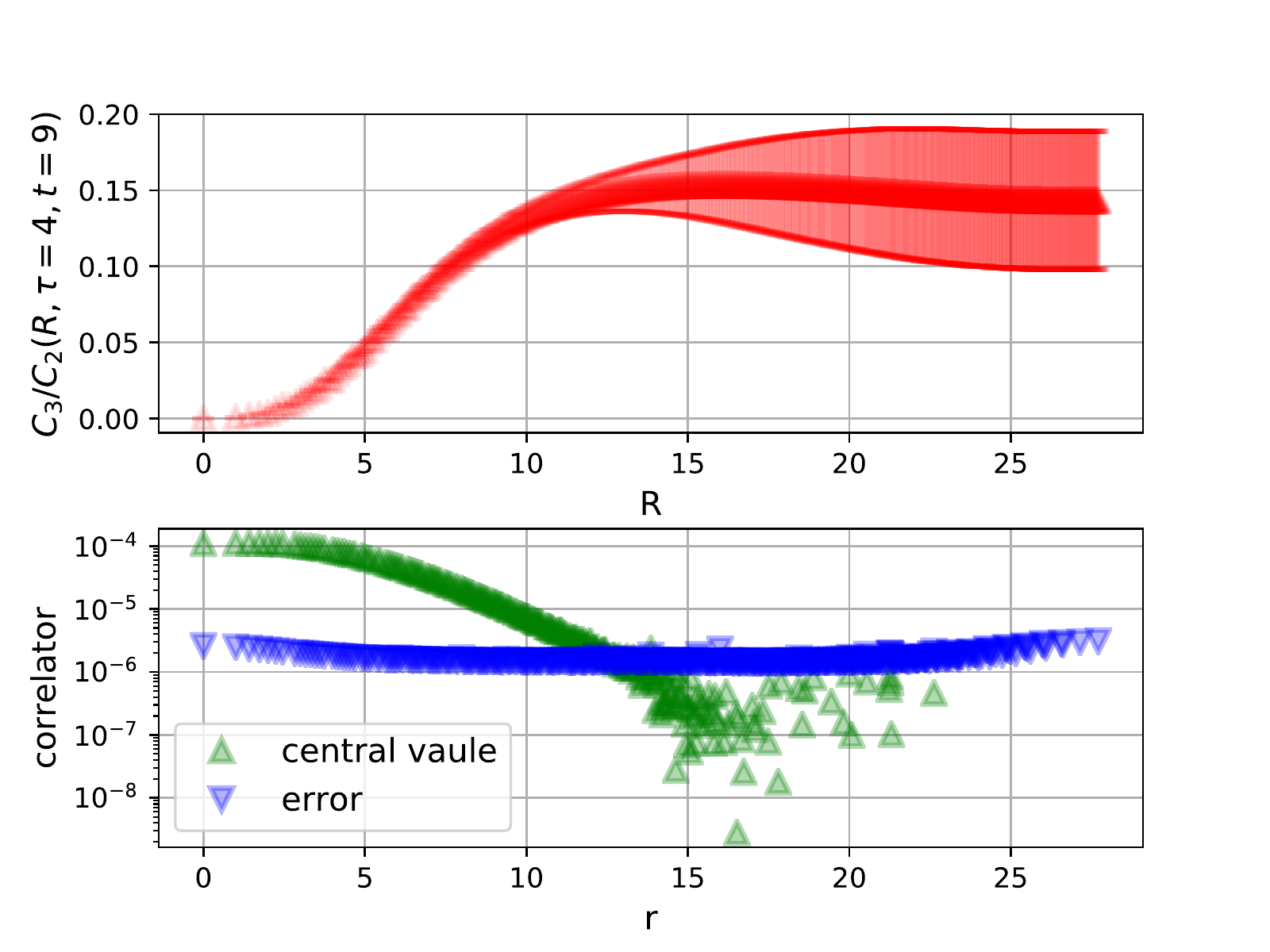}
\caption{The value and error of $C_3/C_2 (R, \tau=5,t=10)$ are plotted in the upper panel as a function of $R$. 
The lower panel displays the three-point function in Eq.~(\ref{eq_dis_ins}) as a function of $r$ without summing over it.
The green band shows the signal and the blue band the error.}
\label{CDP}
\end{figure}
\begin{figure}[t]
\centering
\includegraphics[width=0.4\textwidth]{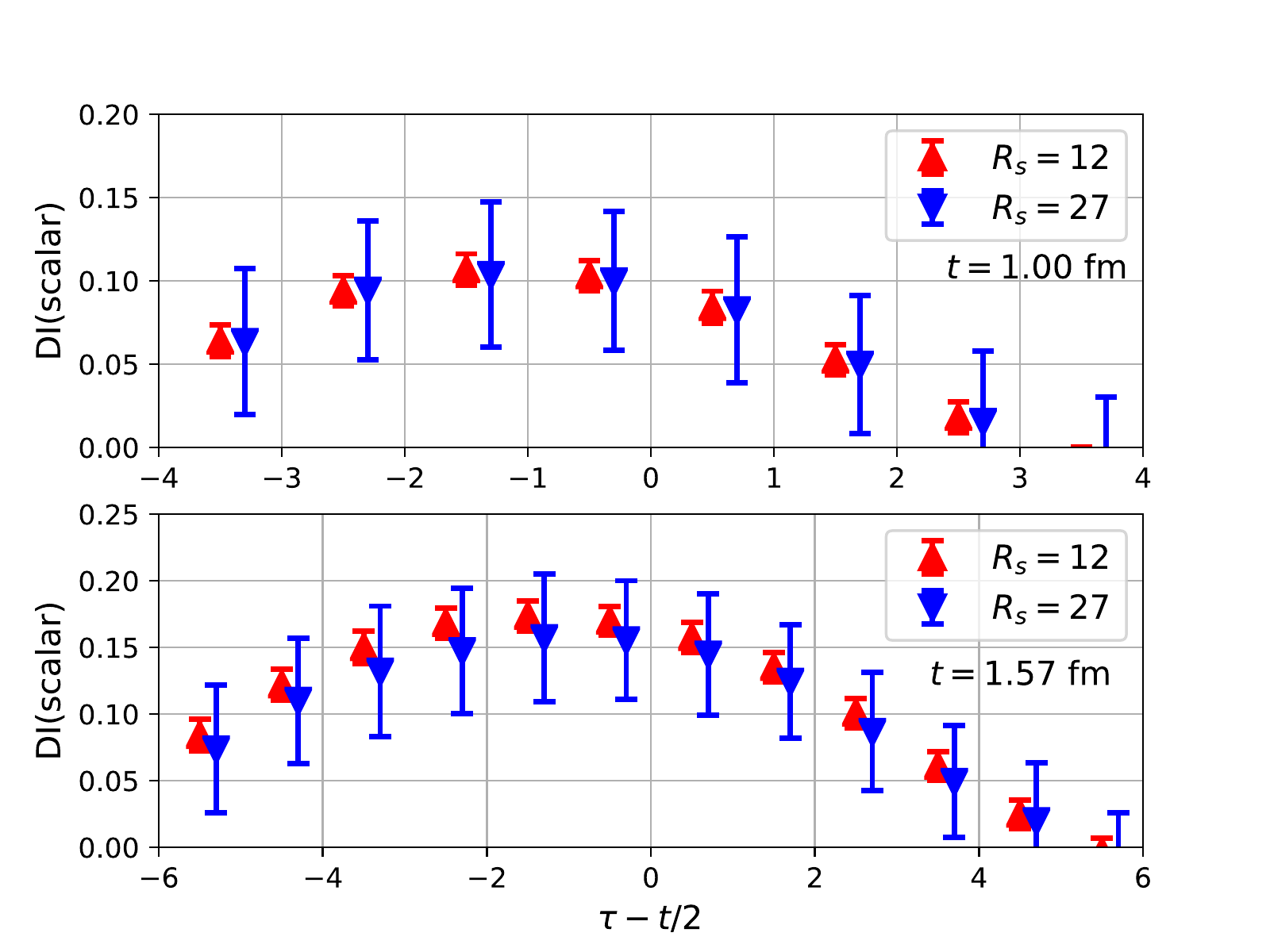}
\caption{DI calculation for the strange scalar matrix element in the nucleon  as a function of
$\tau -t/2$. For each of the source-sink separations at 1.00 fm (upper panel) and 1.57 fm (lower panel),
two results with a cutoff of  $R_s=27$ and $R_s=12$ are plotted.}
\label{fig_scalar1}
\end{figure}
The upper panel of Fig.~\ref{CDP} gives the value and the error of the ratio of three-to-two point functions $C_3/C_2(R,\tau,t)$ in Eq.~(\ref{eq_dis_ins}) as a function of $R$ at 
$\tau = 4$ and $t =9 $. The nucleon source-sink separation is $1.29$ fm in this case. We see the error grows after $R$ is greater than $\sim 12\, (\sim 1.7 {\rm fm})$ while the central value remains constant within errors. This behavior reflects the fact that the signal without summing over $|\vec{r'}|$ falls off exponentially with $r$, while the error remains constant as shown in the lower panel.  


Fig.~\ref{fig_scalar1} shows $DI (scalar)$-- the disconnected three-point to two-point function ratio to obtain the scalar matrix element for the strange quark in the nucleon as a function of $\tau -t/2$ for two source-sink separations at 1.00 fm (upper panel) and 1.57 fm (lower panel). Two results with cutoffs of $R_s = 27$ and $R_s = 12$ are plotted. $R_s$ is the cutoff radius for the relative coordinate between the sink and the quark loop in the spatial sum. It can be seen that the central values of the two cutoffs are all consistent within errors, while the errors with cutoff $R_s = 12$ are smaller than the ones with cutoff $27$, which includes the whole spatial volume,  by a factor of 4 or so. Thus, cutting off the spatial sum at the saturation distance is equivalent to gaining $\sim 16$ times more statistics.

\subsection{Glueball mass}
\begin{figure}[t]
\centering
\includegraphics[width=0.4\textwidth]{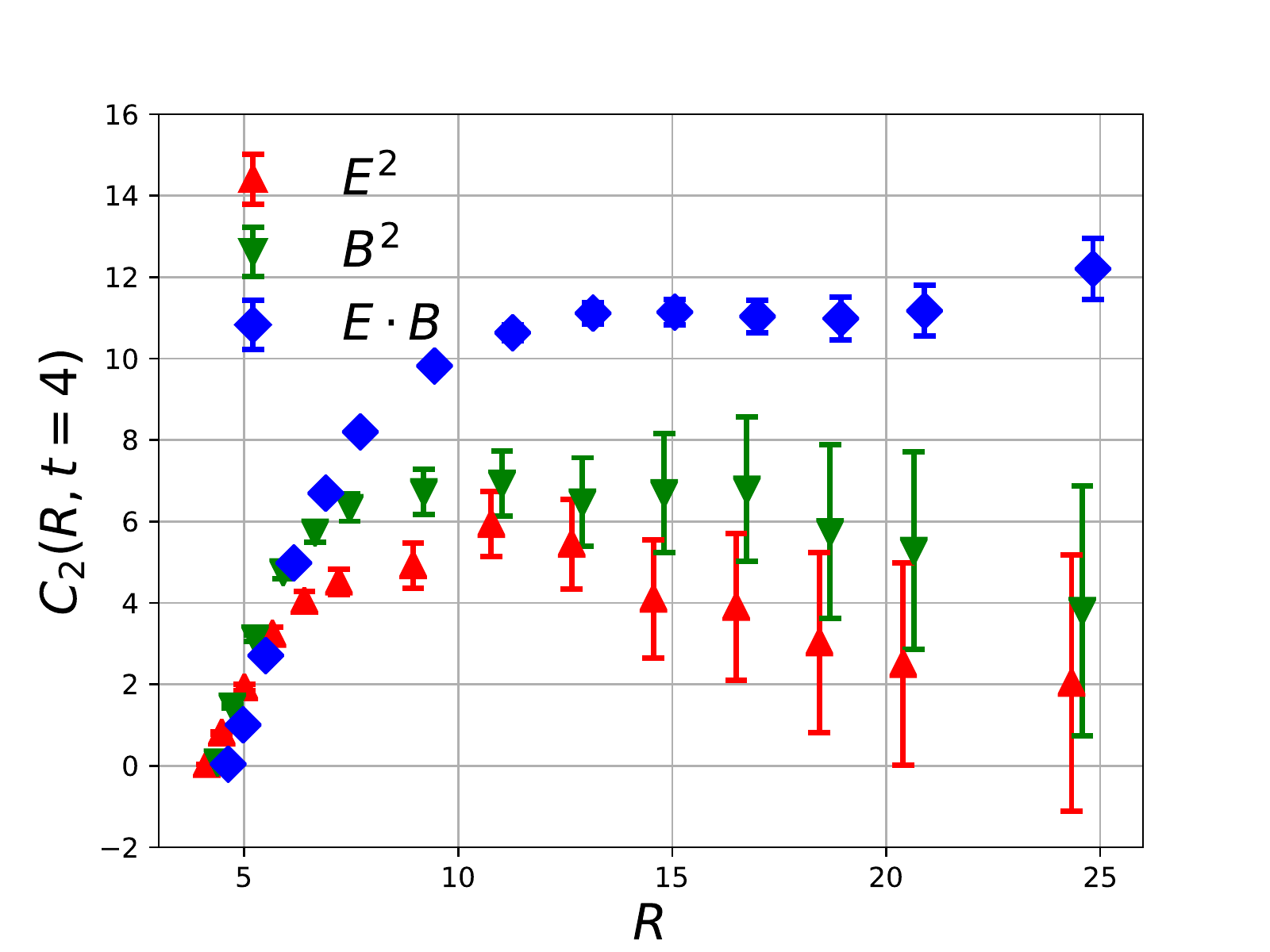}
\caption{Scalar (operator $E^2$ and $B^2$) and pseudo scalar ($E\cdot B$) glueball correlators at $t=4$ as a function of cutoff $R$.}
\label{fig_glueball}
\end{figure}

Next, we consider the glueball correlators in Eq.~(\ref{general_form}) on the 48I lattice with $La = 5.5$ fm.
The correlators from the scalar $E^2$ and $B^2$ operators and the pseudoscalar $E\cdot B$ operator are presented in Fig.~\ref{fig_glueball}, where  they are plotted as a function of $R$ in Eq.~(\ref{general_form}) at $t=4$. We note the scalar correlators saturate after $R\sim9\, (\sim 1.0 {\rm fm})$ and the pseudo scalar one saturates after $R\sim12\, (\sim 1.2 {\rm fm})$,
which can be understood in terms of the different ground state masses in these two channels.
Again, comparing the error at $R=9$ to that at $R = 24$, the latter includes the whole spatial volume, for the scalar case, the error is 
reduced by a factor of $\sim 4$ 
which is in reasonable agreement with the prediction of  $\sim(\frac{24}{9})^\frac{3}{2} = 4.4 $ from Eq.~(\ref{S/N_ratio}). For the pseudoscalar case, the improvement is around 3 times and is consistent with the estimate of $\sim(\frac{24}{12})^\frac{3}{2}=2.8$.

\begin{figure}[b]
\centering
\includegraphics[width=0.4\textwidth]{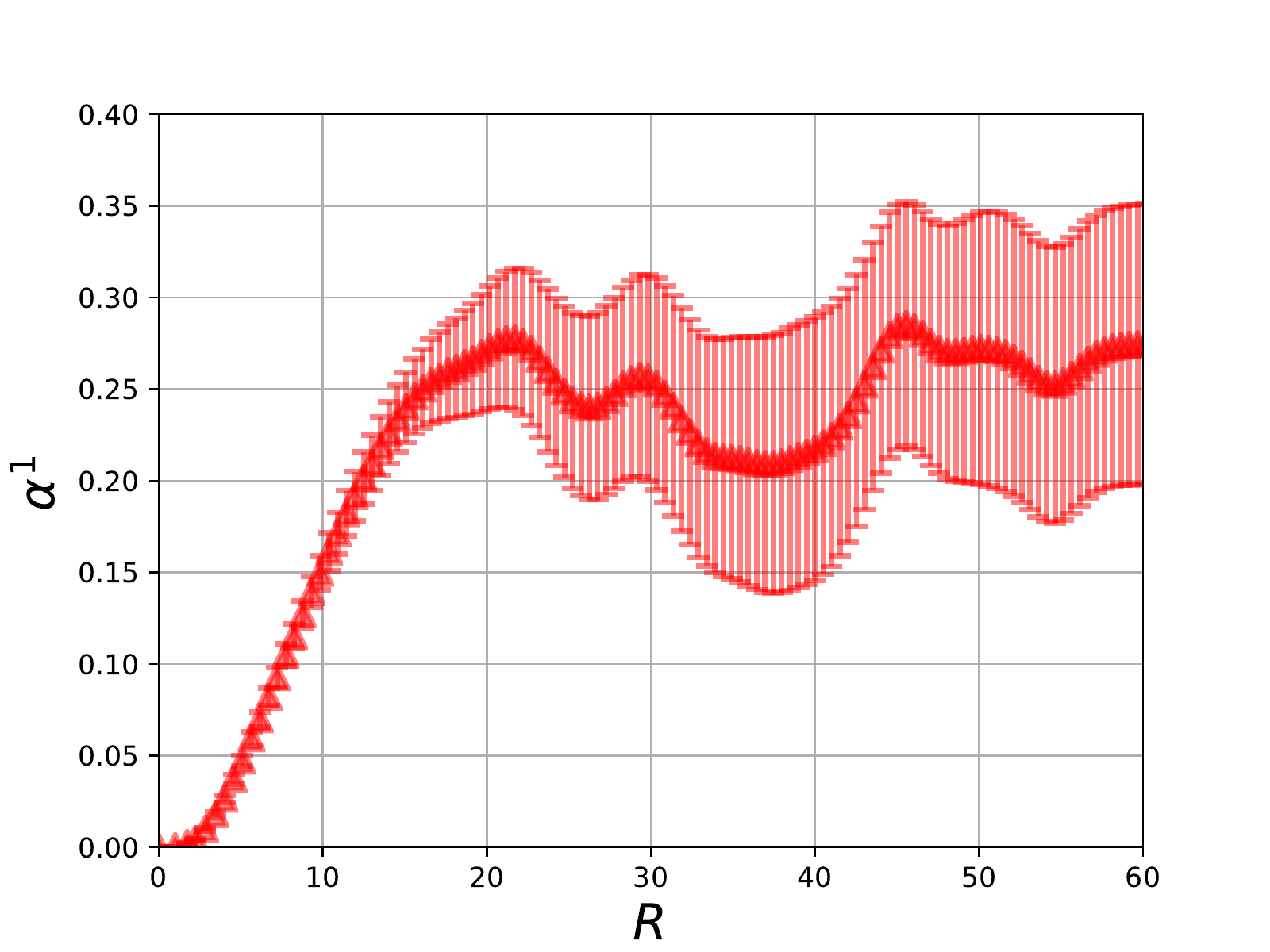}
\caption{The CP-violation phase $\alpha^1$ calculated on the 48I lattice as a function of cutoff $R$.
For each $R$, the value is averaged from $t=6$ to 13.}
\label{fig_alpah1}
\end{figure}
\subsection{Neutron electric dipole moment:}
Finally, we examine the CP-violation phase $\alpha^1$ on the same 48I lattice which is needed for calculating the neutron electric dipole moment (nEDM). The phase is defined as 
\begin{equation}
\alpha^1=\frac{{\rm Tr}[C_{3\mathcal{Q}}(t)\gamma_5]}{{\rm Tr}[C_2(t)\Gamma_e]}
\end{equation}
for large enough $t$, where $C_2(t)$ is the common nucleon two-point function, $\Gamma_e=\frac{1+\gamma_4}{2}$ is the 
parity projector, $C_{3\mathcal{Q}}(t)$ is the nucleon propagator weighted with the total topological charge 
$\mathcal{Q}$
\begin{equation}  \label{Q}
C_{3\mathcal{Q}}(t)=\langle \sum_{\vec{x}}\mathcal{O}_N(\vec{x},t) \bar{\mathcal{O}}_N(\mathcal{G},0)\mathcal{Q}\rangle.
\end{equation}
%
We can turn the total topological charge into the summation of its density, i.e. $\mathcal{Q }= \sum_x q(x)$ where we use the plaquette 
definition for $q(x)$. 
Then the expression of $C_{3\mathcal{Q}}(t)$ with a cutoff $R$ can be cast in the same form as in Eq.~(\ref{eq_dis_ins}), 
 except the scalar quark quark loop $S$ is replaced with the local topological charge $q(x)$ and the sum of the topological charge density  is over the four sphere with a radius $R$.


The result  of $\alpha^1$ as a function of $R$ in Fig.~\ref{fig_alpah1} shows that the signal saturates after $R\sim16$. Cutting off
the sum of $q(x)$ at this R leads to a factor of  $~\sim 3.6$  times reduction in error compared to the case of reweighting with
the total topological charge as in Eq.~(\ref{Q}). This example indicates that for four-dimensional sums, our new method employing the CDP can also improve the S/N.  As we illustrated in the introduction, the nEDM from the $\theta$ term suffers from a $\sqrt{V}$ problem. It is shown here it is related to the vacuum insertion in the variance. This problem is resolved by turning the topological charge into a 4-D sum of the local charge density and applying the CDP by cutting off the relative 4-D distance in the sum. 

  
\section{Systematic and statistical errors}
So far, we have taken a simple cutoff $R_s = 8/M$ to illustrate the efficacy of the variance reduction. This ad hoc choice inevitably 
incurs a systematic error. Since the asymptotic behavior of the integral of the correlator as a function of the separation $R$ is similar to that of the effective mass, we fit it as such and apply the Akaike information criterion (AIC)~\cite{akaike74} to obtain the statistical error and the estimated systematic error by using analysis with different fitting windows and models. The details of the application of AIC to the strange matrix element is provided in the Appendix as an example. It turns out that both the systematic and statistical errors are stable against multiple choices of windows and two fitting models. The statistical error is close to that at the cutoff distance when the plateau emerges (i.e. $R_s = 8/M$). Using a representative fit with 2 fitting formulas and 80 combinations of 8 data points each for a total of 160 fits and 100 bootstrap samplings, we obtain the value of the strange matrix element that we considered earlier to be \mbox{0.160 (15) (15).} The statistical error (first one) and the systematic error (second one) are practically the same. This is to be compared with the original value of 0.143(45) without taking the CDP into account. The systematic error is to be added to the total systematic error of the calculation. 

\section{Discussion and summary}
 Regarding the nucleon correlator in Fig.~\ref{fig_twopoint}, we notice that there is no conspicuous increase of the error as a function of $R$. This is because, unlike the DI, the variance does not have an vacuum insertion for the CI. The leading contribution to
the variance is expected to be $e^{- 3 m_{\pi} r}$. This has a longer range than that of the signal which falls off with the nucleon mass. Therefore, in principle, one would expect some gain in the S/N when $R$ is cut off at $R_s$. Therefore, the corresponding ratio of
S/N in Eq. (\ref{S/N_ratio}) is 
\begin{equation} \label{S/N_ratio_CI}
\frac{S/N(R_s)}{S/N(L)} \sim \sqrt{\frac{A(L/2, 3m_{\pi})}{A(R_s, 3m_{\pi})}}.
\end{equation}
For the 48I lattice in Fig.~\ref{fig_twopoint}, this ratio is 1.13 for the physical pion mass with the cutoff $R_s = 8/M$.
This is not nearly as much a gain as in the DI where the variance is dominated by the vacuum insertion.  In the CI case,
the noise saturates at $\sim 8/(3 m_{\pi})= 3.5$ fm. There is no gain for a lattice with a size larger than this.

In summary, we have shown that the exponential fall off of the Cluster Decomposition Principle (CDP) seems to hold numerically for
the several correlators that we examined. For the disconnected insertions (DI), we find that the vacuum insertion dominates the variance
so that the relevant operators fluctuate independently and is independent of the time separation. This explains why the signal fall off exponentially, while the error remains constant in the DI. To demonstrate the efficacy of employing the CDP to reduce the variance, we have restricted the volume sum of the relative coordinate between the operators to the saturation radius $R_s$ to show that there is an effective gain of $V/V_s$ in statistics without compromising the signal. This applies to all DI cases. For the cases we have considered, namely the glueball mass, the strangeness content in the nucleon, and the CP violation angle in the nucleon due to the $\theta$ term, we found that for lattices with a physical sizes of 4.5 - 5.5 fm, the errors of these quantities can be reduced a factor of  3 to 4. We have applied the Akaike information criterion (AIC)~\cite{akaike74} in the Appendix to estimate the systematic and statistical errors incurred by applying the CDP.  For the strangeness content, we find that the systematic error is practically of
the same size as that of the statistical one when the cluster decomposition is taken into account. This results in a 2 to 3 times reduction in the overall error. For the connected insertions, there is no vacuum insertion in the variance, the gain in statistics is limited. 
 
 \section {Acknowledgement}
 This work is supported in part by the U.S. DOE Grant $\text{No.}$ $\text{DE-SC}0013065$. This research used resources of the Oak Ridge Leadership Computing Facility at the Oak Ridge National Laboratory, which is supported by the Office of Science of the U.S. Department of Energy under Contract No. DE-AC05-00OR22725. This work used the Extreme Science and Engineering Discovery Environment (XSEDE) Stampede at TACC through allocation MCA01S007, which is supported by National Science Foundation grant number ACI-1053575~\cite{XSEDE}. We also thank National Energy Research Scientific Computing Center (NERSC) for providing HPC resources that have contributed to the research results reported within this paper. We acknowledge the facilities of the USQCD Collaboration used for this research in part, which are funded by the Office of Science of the U.S. Department of Energy.

\section{Appendix} 

\subsection{Error estimate using Akaike information criterion (AIC)}

AIC is founded in information theory. It is an estimator of the relative relevance of the statistical models that are used to
describe a given set of data~\cite{akaike74,AIC}. \\

 The usual task is to fit the data generated by some unknown process (function) $f$ with different trial models.   
 However, we cannot tell which model is a better representation of $f$ with certainty in practice, because we do not know $f$.
{Akaike (1974)} showed that we can estimate, via AIC, how much more (or less)
information is lost when comparing one model to another. The estimate, though, is only valid asymptotically. If the number of data 
points is small, then some correction is often necessary (e.g., AICc)~\cite{AIC}. 

\subsection{Definition of the AIC value}

\hspace{0.4cm}
Considering a model $M$ for some data $x$ with $k$ parameters, the
maximum value of the likelihood function for the model $L_{\max}$ is represented by
the following probability
\begin{equation}
  L_{\max} = P (x | \vec{\theta}, M),
\end{equation}
where $\vec{\theta}$ is the parameter vector that maximizes the likelihood
function. The maximum likelihood function is related to the minimum $\chi^2$ values in the
standard fitting via
\begin{equation}
  L_{\max} = e^{- \frac{\chi_{\min}^2}{2}} .
\end{equation}
\hspace{0.4cm}
The AIC value of the model is defined as (see ref.~\cite{AIC} and the reference therein for details)
\begin{equation}
  {\rm AIC} = 2 k - 2 \ln (L_{\max}) = 2 k + \chi^2_{\min},
\end{equation}
This favors models with the minimum AIC value and  penalizes those with many fitting parameters.

\subsection{Practical application}

\hspace{0.4cm} 
In practice, we take a weighted average of all
the models as is carried out in Ref.~\cite{Borsanyi:2014jba,Durr:2015dna}. The normalized weight for each model is
\begin{equation}   \label{weight}
  w_{i =} \frac{e^{- \frac{{\rm AIC}_i}{2}}}{\sum_i e^{-  \frac{{\rm AIC}_i}{2}}},
\end{equation}
where $i$ is the index of the models. \\

When handling the systematic errors of lattice calculations, we usually need
to take consideration of various fitting formulas with different combinations
of data points used for the fit. The AIC method can be helpful in these cases.
Let's consider a case where we have $N$ configurations and, for each configuration, there are $M$ data
points (e.g. different time separation $t$ of a hadron two-point correlator). We plan to use $P$ models
to fit the data (the number $P$ includes different fit ranges and different
combinations of data points) in order to obtain the mean value, the
systematic error and the statistical error of some model parameters (e.g. the
mass of the ground state). The detailed procedure are given as follows: 

\begin{enumerate} 
\item The mean value is the weighted average of all the $P$ fit models (formulas
and combinations of data points).
\begin{equation}
  \bar{x} = \sum_{i = 1}^P w_i x_i,
\end{equation}
$x_i$ is the fitting result from each model and $w_i$ is the normalized weight in Eq.~(\ref{weight}).

\item The systematic error is taken to be the standard error of the weighted mean,
\begin{equation}
  \sigma_{\rm sys} = \sqrt{\sum_{i = 1}^P w_i \sigma_i^2}, \hspace{1cm}
  \sigma_i = (x_i - \bar{x}).
\end{equation}

\item The statistical error can be obtained from bootstrap resampling. We first
do $N_b$ times bootstrap operation, in each bootstrap sample, we fit these $P$
models and save the weighted mean value. After that, we will have $N_b$ AIC
weighted mean values. The bootstrap error of these weighted mean values gives
the final statistical error.

\end{enumerate}

In summary, we need to do $ (1 + N_b) \times P$ times of correlated fits to
obtain all the relevant results.

\subsection{The cluster decomposition case}

In the cluster decomposition case, we shall consider the strange matrix element as a function of the cutoff 
radius $R$ as illustrated in the following figure. We can use the AIC method to estimate the mean
value, the systematic error and the statistical error of the ratio.

To apply the AIC method to this particular case, we need first to determine our
fitting formulas. In view of the fact that the ratio between the three-point function and the two-point function falls off exponentially as a function
of the relative separation between the quark loop and the sink of the nucleon propagator, the accumulated sum of the separation with a cutoff $R$ is expected to be a constant after certain $R$, such as $R_s = 8/M$, as illustrated in the upper panel of Fig.~\ref{CDP}. This is much like fitting the effective mass plot to isolate the ground state. The two formulas (models) we will use are
\begin{equation}
  f_1 (R) = C_0
\end{equation}
and
\begin{equation}
  f_2 (R) = C_0 + C_1 \frac{\sqrt{R} e^{- m R}}{m} .
\end{equation}

The first one is the asymptotic form for $R \rightarrow \infty$ and the second
one is the form commensurate with that from the cluster decomposition principle to cover more range of $R$ in the fitting.

Then, we need to choose the combinations of data points. To make sure that
every combination has the same weight, we set the number of data points
(marked as $N_d$) contained in each combination to be equal. And to enlarge
the number of different combinations (marked as $N_c$) we can have, we do not
force the points in each combination to be contiguous. (e.g., if the total
range of data points is $R \in [8, 27]$, and we set $N_d = 4$, $N_c =
5$, the possible combinations can be $[8, 9, 12, 15]$, $[15, 18, 21, 22]$,
$[15, 16, 17, 18]$, $[10, 12, 20, 22]$ and $[20, 22, 24, 27]$.) If $N_c$ is
large enough, combinations will include both contiguous data points and
noncontiguous data point. So in this sense, this is a more general way to
choose data points.

Having the formulas and combinations, we can then proceed to do the fittings.
The final results might be affected by 3 factors: $N_b$ (number of bootstraps),
$N_d$ (number of data points) and $N_c$ (number of combinations). We vary
these 3 factors to check if the results are stable in these fits.

\begin{center}
\begin{table} [ht]
\caption{The mean values, and the systematic and statistical errors are given for various fits. $N_b$ is the number
of bootstrap samples,  $N_d$ the number is data points in each fit, and $N_c$ is the number of different combinations of the
$N_d$ data points.}
\vspace{0.5cm}
  \begin{tabular}{|p{1.0cm}|p{1.0cm}|p{1.0cm}|p{1.5cm}|p{1.5cm}|p{1.5cm}|}
     \hline
    $N_b$ & $N_d$ & $N_c$ & mean & $E_{{\rm sys}}$ & $E_{{\rm sta}}$\\
    \hline
    100 & 6 & 80 & 0.161 & 0.013 & 0017\\
    100 & 8 & 80 & 0.160 & 0.015 & 0.015\\
    100 & 10 & 80 & 0.163 & 0.019 & 0.012\\
    \hline
    100 & 8 & 60 & 0.161 & 0.015 & 0.015\\
    100 & 8 & 80 & 0.160 & 0.015 & 0.015\\
    100 & 8 & 100 & 0.161 & 0.015 & 0.015\\
    \hline
    50 & 8 & 80 & 0.160 & 0.015 & 0.015\\
    100 & 8 & 80 & 0.160 & 0.015 & 0.015\\
    200 & 8 & 80 & 0.160 & 0.015 & 0.015\\
    \hline
  \end{tabular}
  \end{table}
\end{center}

The data range used is $R \in [10, 27]$. After several hundreds of
thousands of correlated fits, the results are show in the above table. The
values and errors are pretty stable no matter how we vary $N_b$, $N_d$ or $N_c$.
We decide to take the representative results from  $N_b=100, N_d = 8$, and $N_c = 80$ which
gives the value of $0.160 (15) (15)$ as our final estimation. The systematic error 
(the second parenthesis) is comparable to the statistical one (the first parenthesis) from 100 bootstrap samples. 
This is to be compared to the original value of 0.143(45) when the sum over the relative coordinate between the quark loop 
and the sink of the nucleon propagator is carried out to cover the whole spatial volume.
The total number of analysis $P$ is the product of the number of models (2 in this case) 
and $N_c$, the number of different combinations of $N_d$ data points. In this case, 
$P = 2 \times 80 = 160$.

\bibliographystyle{apsrev4-1}
\bibliography{library}

\begin{thebibliography}{20}%
\makeatletter
\providecommand \@ifxundefined [1]{%
 \@ifx{#1\undefined}
}%
\providecommand \@ifnum [1]{%
 \ifnum #1\expandafter \@firstoftwo
 \else \expandafter \@secondoftwo
 \fi
}%
\providecommand \@ifx [1]{%
 \ifx #1\expandafter \@firstoftwo
 \else \expandafter \@secondoftwo
 \fi
}%
\providecommand \natexlab [1]{#1}%
\providecommand \enquote  [1]{``#1''}%
\providecommand \bibnamefont  [1]{#1}%
\providecommand \bibfnamefont [1]{#1}%
\providecommand \citenamefont [1]{#1}%
\providecommand \href@noop [0]{\@secondoftwo}%
\providecommand \href [0]{\begingroup \@sanitize@url \@href}%
\providecommand \@href[1]{\@@startlink{#1}\@@href}%
\providecommand \@@href[1]{\endgroup#1\@@endlink}%
\providecommand \@sanitize@url [0]{\catcode `\\12\catcode `\$12\catcode
  `\&12\catcode `\#12\catcode `\^12\catcode `\_12\catcode `\%12\relax}%
\providecommand \@@startlink[1]{}%
\providecommand \@@endlink[0]{}%
\providecommand \url  [0]{\begingroup\@sanitize@url \@url }%
\providecommand \@url [1]{\endgroup\@href {#1}{\urlprefix }}%
\providecommand \urlprefix  [0]{URL }%
\providecommand \Eprint [0]{\href }%
\providecommand \doibase [0]{http://dx.doi.org/}%
\providecommand \selectlanguage [0]{\@gobble}%
\providecommand \bibinfo  [0]{\@secondoftwo}%
\providecommand \bibfield  [0]{\@secondoftwo}%
\providecommand \translation [1]{[#1]}%
\providecommand \BibitemOpen [0]{}%
\providecommand \bibitemStop [0]{}%
\providecommand \bibitemNoStop [0]{.\EOS\space}%
\providecommand \EOS [0]{\spacefactor3000\relax}%
\providecommand \BibitemShut  [1]{\csname bibitem#1\endcsname}%
\let\auto@bib@innerbib\@empty
\bibitem [{\citenamefont {Aoki}\ \emph {et~al.}(2017)\citenamefont {Aoki} \emph
  {et~al.}}]{Aoki:2016frl}%
  \BibitemOpen
  \bibfield  {author} {\bibinfo {author} {\bibfnamefont {S.}~\bibnamefont
  {Aoki}} \emph {et~al.},\ }\href {\doibase 10.1140/epjc/s10052-016-4509-7}
  {\bibfield  {journal} {\bibinfo  {journal} {Eur. Phys. J.}\ }\textbf
  {\bibinfo {volume} {C77}},\ \bibinfo {pages} {112} (\bibinfo {year}
  {2017})},\ \Eprint {http://arxiv.org/abs/1607.00299} {arXiv:1607.00299
  [hep-lat]} \BibitemShut {NoStop}%
\bibitem [{\citenamefont {Parisi}(1984)}]{Parisi:1983ae}%
  \BibitemOpen
  \bibfield  {author} {\bibinfo {author} {\bibfnamefont {G.}~\bibnamefont
  {Parisi}},\ }\bibfield  {booktitle} {\emph {\bibinfo {booktitle} {{COMMON
  TRENDS IN PARTICLE AND CONDENSED MATTER PHYSICS. PROCEEDINGS, WINTER ADVANCED
  STUDY INSTITUTE, LES HOUCHES, FRANCE, FEBRUARY 23 - MARCH 11, 1983}}},\
  }\href {\doibase 10.1016/0370-1573(84)90081-4} {\bibfield  {journal}
  {\bibinfo  {journal} {Phys. Rept.}\ }\textbf {\bibinfo {volume} {103}},\
  \bibinfo {pages} {203} (\bibinfo {year} {1984})}\BibitemShut {NoStop}%
\bibitem [{\citenamefont {Lepage}(1989)}]{Lepage:1989hd}%
  \BibitemOpen
  \bibfield  {author} {\bibinfo {author} {\bibfnamefont {G.~P.}\ \bibnamefont
  {Lepage}},\ }in\ \href {http://alice.cern.ch/format/showfull?sysnb=0117836}
  {\emph {\bibinfo {booktitle} {{Boulder ASI 1989:97-120}}}}\ (\bibinfo {year}
  {1989})\ pp.\ \bibinfo {pages} {97--120}\BibitemShut {NoStop}%
\bibitem [{\citenamefont {Araki}\ \emph {et~al.}(1962)\citenamefont {Araki},
  \citenamefont {Hepp},\ and\ \citenamefont {Ruelle}}]{Araki:1962zhd}%
  \BibitemOpen
  \bibfield  {author} {\bibinfo {author} {\bibfnamefont {H.}~\bibnamefont
  {Araki}}, \bibinfo {author} {\bibfnamefont {K.}~\bibnamefont {Hepp}}, \ and\
  \bibinfo {author} {\bibfnamefont {D.}~\bibnamefont {Ruelle}},\ }\href
  {\doibase 10.5169/seals-113273} {\bibfield  {journal} {\bibinfo  {journal}
  {Helv. Phys. Acta}\ }\textbf {\bibinfo {volume} {35}},\ \bibinfo {pages}
  {164} (\bibinfo {year} {1962})}\BibitemShut {NoStop}%
\bibitem [{\citenamefont {Bali}\ \emph {et~al.}(2009)\citenamefont {Bali},
  \citenamefont {Collins},\ and\ \citenamefont {Schafer}}]{Bali:2009dz}%
  \BibitemOpen
  \bibfield  {author} {\bibinfo {author} {\bibfnamefont {G.}~\bibnamefont
  {Bali}}, \bibinfo {author} {\bibfnamefont {S.}~\bibnamefont {Collins}}, \
  and\ \bibinfo {author} {\bibfnamefont {A.}~\bibnamefont {Schafer}} (\bibinfo
  {collaboration} {QCDSF}),\ }\href@noop {} {\bibfield  {journal} {\bibinfo
  {journal} {PoS}\ }\textbf {\bibinfo {volume} {LAT2009}},\ \bibinfo {pages}
  {149} (\bibinfo {year} {2009})},\ \Eprint {http://arxiv.org/abs/0911.2407}
  {arXiv:0911.2407 [hep-lat]} \BibitemShut {NoStop}%
\bibitem [{\citenamefont {Chen}\ \emph {et~al.}(2015)\citenamefont {Chen},
  \citenamefont {Alexandru}, \citenamefont {Draper}, \citenamefont {Liu},
  \citenamefont {Liu},\ and\ \citenamefont {Yang}}]{Chen:2015tpa}%
  \BibitemOpen
  \bibfield  {author} {\bibinfo {author} {\bibfnamefont {Y.}~\bibnamefont
  {Chen}}, \bibinfo {author} {\bibfnamefont {A.}~\bibnamefont {Alexandru}},
  \bibinfo {author} {\bibfnamefont {T.}~\bibnamefont {Draper}}, \bibinfo
  {author} {\bibfnamefont {K.-F.}\ \bibnamefont {Liu}}, \bibinfo {author}
  {\bibfnamefont {Z.}~\bibnamefont {Liu}}, \ and\ \bibinfo {author}
  {\bibfnamefont {Y.-B.}\ \bibnamefont {Yang}},\ }\href@noop {} {\  (\bibinfo
  {year} {2015})},\ \Eprint {http://arxiv.org/abs/1507.02541} {arXiv:1507.02541
  [hep-ph]} \BibitemShut {NoStop}%
\bibitem [{\citenamefont {Blum}\ \emph
  {et~al.}(2016{\natexlab{a}})\citenamefont {Blum}, \citenamefont {Christ},
  \citenamefont {Hayakawa}, \citenamefont {Izubuchi}, \citenamefont {Jin},\
  and\ \citenamefont {Lehner}}]{Blum:2015gfa}%
  \BibitemOpen
  \bibfield  {author} {\bibinfo {author} {\bibfnamefont {T.}~\bibnamefont
  {Blum}}, \bibinfo {author} {\bibfnamefont {N.}~\bibnamefont {Christ}},
  \bibinfo {author} {\bibfnamefont {M.}~\bibnamefont {Hayakawa}}, \bibinfo
  {author} {\bibfnamefont {T.}~\bibnamefont {Izubuchi}}, \bibinfo {author}
  {\bibfnamefont {L.}~\bibnamefont {Jin}}, \ and\ \bibinfo {author}
  {\bibfnamefont {C.}~\bibnamefont {Lehner}},\ }\href {\doibase
  10.1103/PhysRevD.93.014503} {\bibfield  {journal} {\bibinfo  {journal} {Phys.
  Rev.}\ }\textbf {\bibinfo {volume} {D93}},\ \bibinfo {pages} {014503}
  (\bibinfo {year} {2016}{\natexlab{a}})},\ \Eprint
  {http://arxiv.org/abs/1510.07100} {arXiv:1510.07100 [hep-lat]} \BibitemShut
  {NoStop}%
\bibitem [{\citenamefont {C{\`e}}\ \emph {et~al.}(2017)\citenamefont {C{\`e}},
  \citenamefont {Giusti},\ and\ \citenamefont {Schaefer}}]{Ce:2016ajy}%
  \BibitemOpen
  \bibfield  {author} {\bibinfo {author} {\bibfnamefont {M.}~\bibnamefont
  {C{\`e}}}, \bibinfo {author} {\bibfnamefont {L.}~\bibnamefont {Giusti}}, \
  and\ \bibinfo {author} {\bibfnamefont {S.}~\bibnamefont {Schaefer}},\ }\href
  {\doibase 10.1103/PhysRevD.95.034503} {\bibfield  {journal} {\bibinfo
  {journal} {Phys. Rev.}\ }\textbf {\bibinfo {volume} {D95}},\ \bibinfo {pages}
  {034503} (\bibinfo {year} {2017})},\ \Eprint
  {http://arxiv.org/abs/1609.02419} {arXiv:1609.02419 [hep-lat]} \BibitemShut
  {NoStop}%
\bibitem [{\citenamefont {Shintani}\ \emph {et~al.}(2016)\citenamefont
  {Shintani}, \citenamefont {Blum}, \citenamefont {Izubuchi},\ and\
  \citenamefont {Soni}}]{Shintani:2015vsx}%
  \BibitemOpen
  \bibfield  {author} {\bibinfo {author} {\bibfnamefont {E.}~\bibnamefont
  {Shintani}}, \bibinfo {author} {\bibfnamefont {T.}~\bibnamefont {Blum}},
  \bibinfo {author} {\bibfnamefont {T.}~\bibnamefont {Izubuchi}}, \ and\
  \bibinfo {author} {\bibfnamefont {A.}~\bibnamefont {Soni}},\ }\href {\doibase
  10.1103/PhysRevD.93.094503} {\bibfield  {journal} {\bibinfo  {journal} {Phys.
  Rev.}\ }\textbf {\bibinfo {volume} {D93}},\ \bibinfo {pages} {094503}
  (\bibinfo {year} {2016})},\ \Eprint {http://arxiv.org/abs/1512.00566}
  {arXiv:1512.00566 [hep-lat]} \BibitemShut {NoStop}%
\bibitem [{\citenamefont {Li}\ \emph {et~al.}(2010)\citenamefont {Li} \emph
  {et~al.}}]{Li:2010pw}%
  \BibitemOpen
  \bibfield  {author} {\bibinfo {author} {\bibfnamefont {A.}~\bibnamefont {Li}}
  \emph {et~al.} (\bibinfo {collaboration} {xQCD}),\ }\href {\doibase
  10.1103/PhysRevD.82.114501} {\bibfield  {journal} {\bibinfo  {journal} {Phys.
  Rev.}\ }\textbf {\bibinfo {volume} {D82}},\ \bibinfo {pages} {114501}
  (\bibinfo {year} {2010})},\ \Eprint {http://arxiv.org/abs/1005.5424}
  {arXiv:1005.5424 [hep-lat]} \BibitemShut {NoStop}%
\bibitem [{\citenamefont {Gong}\ \emph {et~al.}(2013)\citenamefont {Gong} \emph
  {et~al.}}]{Gong:2013vja}%
  \BibitemOpen
  \bibfield  {author} {\bibinfo {author} {\bibfnamefont {M.}~\bibnamefont
  {Gong}} \emph {et~al.} (\bibinfo {collaboration} {XQCD}),\ }\href {\doibase
  10.1103/PhysRevD.88.014503} {\bibfield  {journal} {\bibinfo  {journal} {Phys.
  Rev.}\ }\textbf {\bibinfo {volume} {D88}},\ \bibinfo {pages} {014503}
  (\bibinfo {year} {2013})},\ \Eprint {http://arxiv.org/abs/1304.1194}
  {arXiv:1304.1194 [hep-ph]} \BibitemShut {NoStop}%
\bibitem [{\citenamefont {Sufian}\ \emph {et~al.}(2017)\citenamefont {Sufian},
  \citenamefont {Yang}, \citenamefont {Alexandru}, \citenamefont {Draper},
  \citenamefont {Liang},\ and\ \citenamefont {Liu}}]{Sufian:2016pex}%
  \BibitemOpen
  \bibfield  {author} {\bibinfo {author} {\bibfnamefont {R.~S.}\ \bibnamefont
  {Sufian}}, \bibinfo {author} {\bibfnamefont {Y.-B.}\ \bibnamefont {Yang}},
  \bibinfo {author} {\bibfnamefont {A.}~\bibnamefont {Alexandru}}, \bibinfo
  {author} {\bibfnamefont {T.}~\bibnamefont {Draper}}, \bibinfo {author}
  {\bibfnamefont {J.}~\bibnamefont {Liang}}, \ and\ \bibinfo {author}
  {\bibfnamefont {K.-F.}\ \bibnamefont {Liu}},\ }\href {\doibase
  10.1103/PhysRevLett.118.042001} {\bibfield  {journal} {\bibinfo  {journal}
  {Phys. Rev. Lett.}\ }\textbf {\bibinfo {volume} {118}},\ \bibinfo {pages}
  {042001} (\bibinfo {year} {2017})},\ \Eprint
  {http://arxiv.org/abs/1606.07075} {arXiv:1606.07075 [hep-ph]} \BibitemShut
  {NoStop}%
\bibitem [{\citenamefont {Yang}\ \emph
  {et~al.}(2016{\natexlab{a}})\citenamefont {Yang}, \citenamefont {Alexandru},
  \citenamefont {Draper}, \citenamefont {Gong},\ and\ \citenamefont
  {Liu}}]{Yang:2015zja}%
  \BibitemOpen
  \bibfield  {author} {\bibinfo {author} {\bibfnamefont {Y.-B.}\ \bibnamefont
  {Yang}}, \bibinfo {author} {\bibfnamefont {A.}~\bibnamefont {Alexandru}},
  \bibinfo {author} {\bibfnamefont {T.}~\bibnamefont {Draper}}, \bibinfo
  {author} {\bibfnamefont {M.}~\bibnamefont {Gong}}, \ and\ \bibinfo {author}
  {\bibfnamefont {K.-F.}\ \bibnamefont {Liu}},\ }\href {\doibase
  10.1103/PhysRevD.93.034503} {\bibfield  {journal} {\bibinfo  {journal} {Phys.
  Rev.}\ }\textbf {\bibinfo {volume} {D93}},\ \bibinfo {pages} {034503}
  (\bibinfo {year} {2016}{\natexlab{a}})},\ \Eprint
  {http://arxiv.org/abs/1509.04616} {arXiv:1509.04616 [hep-lat]} \BibitemShut
  {NoStop}%
\bibitem [{\citenamefont {Blum}\ \emph
  {et~al.}(2016{\natexlab{b}})\citenamefont {Blum} \emph
  {et~al.}}]{Blum:2014tka}%
  \BibitemOpen
  \bibfield  {author} {\bibinfo {author} {\bibfnamefont {T.}~\bibnamefont
  {Blum}} \emph {et~al.} (\bibinfo {collaboration} {RBC, UKQCD}),\ }\href
  {\doibase 10.1103/PhysRevD.93.074505} {\bibfield  {journal} {\bibinfo
  {journal} {Phys. Rev.}\ }\textbf {\bibinfo {volume} {D93}},\ \bibinfo {pages}
  {074505} (\bibinfo {year} {2016}{\natexlab{b}})},\ \Eprint
  {http://arxiv.org/abs/1411.7017} {arXiv:1411.7017 [hep-lat]} \BibitemShut
  {NoStop}%
\bibitem [{\citenamefont {Yang}\ \emph
  {et~al.}(2016{\natexlab{b}})\citenamefont {Yang}, \citenamefont {Alexandru},
  \citenamefont {Draper}, \citenamefont {Liang},\ and\ \citenamefont
  {Liu}}]{Yang:2015uis}%
  \BibitemOpen
  \bibfield  {author} {\bibinfo {author} {\bibfnamefont {Y.-B.}\ \bibnamefont
  {Yang}}, \bibinfo {author} {\bibfnamefont {A.}~\bibnamefont {Alexandru}},
  \bibinfo {author} {\bibfnamefont {T.}~\bibnamefont {Draper}}, \bibinfo
  {author} {\bibfnamefont {J.}~\bibnamefont {Liang}}, \ and\ \bibinfo {author}
  {\bibfnamefont {K.-F.}\ \bibnamefont {Liu}} (\bibinfo {collaboration}
  {xQCD}),\ }\href {\doibase 10.1103/PhysRevD.94.054503} {\bibfield  {journal}
  {\bibinfo  {journal} {Phys. Rev.}\ }\textbf {\bibinfo {volume} {D94}},\
  \bibinfo {pages} {054503} (\bibinfo {year} {2016}{\natexlab{b}})},\ \Eprint
  {http://arxiv.org/abs/1511.09089} {arXiv:1511.09089 [hep-lat]} \BibitemShut
  {NoStop}%
\bibitem [{\citenamefont {Akaike}(1974)}]{akaike74}%
  \BibitemOpen
  \bibfield  {author} {\bibinfo {author} {\bibfnamefont {H.}~\bibnamefont
  {Akaike}},\ }\href@noop {} {\bibfield  {journal} {\bibinfo  {journal} {IEEE
  Transactions on Automatic Control}\ }\textbf {\bibinfo {volume} {19}},\
  \bibinfo {pages} {716} (\bibinfo {year} {1974})}\BibitemShut {NoStop}%
\bibitem [{\citenamefont {Towns}\ \emph {et~al.}(2014)\citenamefont {Towns}
  \emph {et~al.}}]{XSEDE}%
  \BibitemOpen
  \bibfield  {author} {\bibinfo {author} {\bibfnamefont {J.}~\bibnamefont
  {Towns}} \emph {et~al.},\ }\href@noop {} {\bibfield  {journal} {\bibinfo
  {journal} {Computing in Science \& Engineering}\ }\textbf {\bibinfo {volume}
  {16}},\ \bibinfo {pages} {62} (\bibinfo {year} {2014})}\BibitemShut {NoStop}%
\bibitem [{AIC()}]{AIC}%
  \BibitemOpen
  \href@noop {} {\bibinfo  {journal}
  {https://en.wikipedia.org/wiki/Akaike\_information\_criterion}\ }\BibitemShut
  {NoStop}%
\bibitem [{\citenamefont {Borsanyi}\ \emph {et~al.}(2015)\citenamefont
  {Borsanyi} \emph {et~al.}}]{Borsanyi:2014jba}%
  \BibitemOpen
\bibfield  {journal} {  }\bibfield  {author} {\bibinfo {author} {\bibfnamefont
  {S.}~\bibnamefont {Borsanyi}} \emph {et~al.},\ }\href {\doibase
  10.1126/science.1257050} {\bibfield  {journal} {\bibinfo  {journal}
  {Science}\ }\textbf {\bibinfo {volume} {347}},\ \bibinfo {pages} {1452}
  (\bibinfo {year} {2015})},\ \Eprint {http://arxiv.org/abs/1406.4088}
  {arXiv:1406.4088 [hep-lat]} \BibitemShut {NoStop}%
\bibitem [{\citenamefont {Durr}\ \emph {et~al.}(2016)\citenamefont {Durr} \emph
  {et~al.}}]{Durr:2015dna}%
  \BibitemOpen
  \bibfield  {author} {\bibinfo {author} {\bibfnamefont {S.}~\bibnamefont
  {Durr}} \emph {et~al.},\ }\href {\doibase 10.1103/PhysRevLett.116.172001}
  {\bibfield  {journal} {\bibinfo  {journal} {Phys. Rev. Lett.}\ }\textbf
  {\bibinfo {volume} {116}},\ \bibinfo {pages} {172001} (\bibinfo {year}
  {2016})},\ \Eprint {http://arxiv.org/abs/1510.08013} {arXiv:1510.08013
  [hep-lat]} \BibitemShut {NoStop}%
\end{thebibliography}%

\end{document}